\documentclass{article}
\usepackage[utf8]{inputenc}
\usepackage[T1]{fontenc}
\usepackage{graphicx}
\usepackage{grffile}
\usepackage{longtable}
\usepackage{wrapfig}
\usepackage{rotating}
\usepackage[normalem]{ulem}
\usepackage{amsmath}
\usepackage{textcomp}
\usepackage{amssymb}
\usepackage{capt-of}
\usepackage{hyperref}
\usepackage{caption}
\usepackage{subcaption}
\usepackage{mathpazo}
\usepackage[backend=biber, style=authoryear, url=false]{biblatex}
\usepackage{longtable}
\usepackage{booktabs}
\usepackage{enumitem}
\usepackage{placeins}
\usepackage{array}
\usepackage{multirow}

\title{ Using relative weight analysis with residualization to detect relevant nonlinear interaction effects in ordinary and logistic regression}
\author{Maikol Solís\footnote{Corresponding author. Universidad de Costa Rica,
		Escuela de Matemática, Centro de Investigación en Matemática Pura y
		Aplicada. email: \url{maikol.solis@ucr.ac.cr}} \and Carlos
	Pasquier\footnote{Universidad de Costa Rica, Escuela de Matemática, Centro de Matemática Pura y Aplicada. email: \url{carlos.pasquier@ucr.ac.cr}}} 
\date{ }
\begin{document}

\maketitle
\begin{abstract}
 Relative weight analysis is a classic tool for detecting whether  one variable or
	interaction in a model is relevant. In this study, we focus on the
	construction of relative weights for non-linear interactions using restricted
	cubic splines. Our aim is to provide an accessible method to analyze a
	multivariate model and identify one subset with the most representative set of
	variables. Furthermore, we developed a procedure for treating control, fixed, free
	and interaction terms simultaneously in the residual weight analysis. The
	interactions are residualized properly against their main effects to maintain
	their true effects in the model. We tested this method using two simulated
	examples.
\end{abstract}

\section{Introduction}

Traditional fit in ordinary or logistic regressions estimates the best
set of parameters for a dataset. We executed another analysis in tandem, like
hypothesis testing about the nullity of coefficients; estimation of confidence
intervals; check the distribution of the residuals; or checking the variability
of the fit along with the data. These checks are mandatory in modern
statistics.

We can also check their influence on the model depending on the number of
variables in the data. Here, we refer to the capacity of the variables to
significantly impact the output. Non-influential variables should be excluded
from the analysis. This analysis helps simplify the problem, analyze the
results better, and choose better decisions.

General solutions to  this analysis were proposed. Some examples use
principal component analysis, stepwise regression, lasso, ridge, or elastic net
regression \parencite{HastieElements2011}. However, these can detect and remove
non-influential variables, but they hinder the impact of removing or adding this
variable with respect to the other variables.

For a detailed measurement of the influence of each variable, we can mention
measures such as zero-order correlations, standardized regression weights, and
semi-partial correlations \parencite{JohnsonHistory2004}. However, if
multicollinearity exists among the variables, these measures are inadequate.
Other techniques, such as dominance analysis \parencite{BudescuDominance1993}
and residual weight analysis, have arisen to capture the complexity of a model
\parencite{JohnsonHeuristic2000}. The dominance analysis and residual weights
analysis aim to estimate the variable's importance with respect to the output
when multiple correlated predictors exist \parencite{LebretonMonte2004}.

In this paper, we focus on the residual weight analysis originally proposed by
\textcite{JohnsonHeuristic2000}. Using ordinary least squares (OLS) regression,
the technique creates a new set of predictors that are orthogonal
representations of the original ones. In this new space, the importance standard
score for each variable is estimated. Then, the scores return to the original
space of variables through a transformation matrix.

The method OLS regression method uses assumptions such as homoscedasticity,
linearity and residual normality. However, in areas such as sociology or
psychology, the dependent variable is categorical, and thus a logistic approach
is necessary. All the aforementioned assumptions are violated in this case. The
work in \textcite{TonidandelDetermining2010} proposed a solution to this problem
by estimating residual weights for logistic regression. The solution creates a
new orthogonal space and estimates the standardized scores with the standard
deviation of the logistic response variable.

With interactions, the work of \textcite{LeBreton2013} proposed
residualizing the interaction terms to account for only the true effect among
the variables. Thus, they executed a local regression where the
dependent variable is the interaction term and the covariates are the main
effects. Then, the residual error captures the pure effect between variables.

Regarding the type of model we use (OLS or logistic), the mentioned techniques
intrinsically depend on the underlying way we adjust the predictors. The usual
linear regression works for most problems. However, if the data have nonlinear
structures; the linearity assumption is insufficient. Therefore, it can be
extended using a nonlinear smoother.

In this work, we perform a relative weight analysis where the dependent variable
is either continuous or binary using restricted splines to model the predictors.
Even if we have a clear knowledge about the main effects of the model, the
interactions remain an obscure part of the problem, making it difficult to match
two covariates that provide the most possible information. Thus, we add the most
relevant interactions terms to the regression to capture all the information not
seen by the main effects. Our procedure includes all the interactions
among the main effects and then searches for the best subset through a stepwise
selection process. Even if there are some criticisms of the stepwise technique
\parencite{MillerSubset2002,HarrellRegression2015}, we get accurate results.


To build a flexible procedure adjusted to multiple needs, we will allow the use
of control variables, fixed variables, free variables and pairwise interactions.
The control variables are static and remain only as main effects; the fixed and
free variables are used to create interactions. The difference is the former
ones remain in the final model, while the later ones can be removed in the
stepwise process.

The remainder of this paper is organized as follows.
Section~\ref{sec:preeliminaries} presents a review of the preliminaries on
setting regression models with restricted splines. In
Section~\ref{sec:methodology} resides the core of the paper, where we explain
all the details to build a procedure of the Relative Weight Analysis with
residualization. We devote Section~\ref{sec:results} to testing the algorithm
capabilities using two simulated examples. Finally, in
Section~\ref{sec:conclusion}, we present the conclusions with some discussion
about future lines of research.

\section{Preliminaries}
\label{sec:preeliminaries}
In a generalized linear model, we can link a set of input variables
\(\boldsymbol{X}=(X_{1},\ldots,X_{d})\) to an output variable \(Y\) through
function \(C\). We also have a set of \(n\) observations for each
\(d\)-tuple of variables. Formally, we state the model as,

\[
	\mathrm{C}(Y\vert \boldsymbol{X}) = \beta _{0} + \beta _{1} X_1 + \cdots \beta _{d} X_{d} +\varepsilon .
\]
The parameters \(\beta_{0}, \dots , \beta _{d}\), are estimated by an
\emph{iterative reweighed the least-squared} procedure. For a detailed review of
logistic regression; we refer the reader to \textcite{HastieElements2011}.

In this work, we will use two cases for the function \(C(\cdot)\)

\begin{description}
	\item[Gaussian:] The output variable \(Y = Y_{g} \in \mathbb{R}\) and we set the link function as \(\mathrm{C}(Y_{g}\vert \boldsymbol{X}) = Y_{g} \), and the \(\varepsilon\) are independent and normally distributed.

	\item[Binomial:] The output variable is classified  as 0 or 1 (\(Y = Y_{b}=\{0,1\}\)). The link function is defined as \(\mathrm{C}(Y_{b}\vert \boldsymbol{X}) = \mathrm{logit}(\mathbb{P}\left(Y_{b}=1\vert \boldsymbol{X} \right))\), where \(\mathrm{logit}(u) = \log(u /(1 - u))\) is a function defined in \([0,1]\to \mathbb{R}\). The noise variable \(\varepsilon\) has an independent logistic distribution.
\end{description}

In the following sections, and to simplify the notation, we will use the
parameters named as \(\beta \) in different contexts without implying that they
are the same. Each equation has exactly the parameters that we want to present,
unless otherwise stated. Additionally, we will use only \(Y\) to denote either
\(Y_{g}\) or \(Y_{b}\). In numerical examples, we will retake the notation
to differentiate each case.

\subsection{Control, fixed, and free variables}\label{sec:control-fixed-free}

We assume  a generalized linear model with the form,

\begin{equation}
	\label{eq:base-model}
	\mathrm{C}(Y\vert \boldsymbol{X})  =\beta_{0} + \sum_{i=1}^{c} \gamma_{i}Q_{i}
	+ \sum_{i=1}^{d_{F}} f_{i}(X_{i}) + \sum_{i=d_{F}+1}^{p} f_{i}(X_{i}).
\end{equation}

The model is formed by \(Q_1, \dots , Q_c\) control
variables, \(f_1(X_1), \dots , f_{d_{F}}(X_{d_{F}})\) fixed variables and
\(f_{d_{F}+1}(X_{d_{F}+1}), \dots , f_{p}(X_{n})\) free variables.

The functions $f$ represents the smoother used for each variable. In the classic
setting, for a variable \(X\) we put \(f(X) = \beta_{1} X\). However, to allow a
parsimonious structure to the model; we should to introduce nonlinear functions.

We use restricted splines to model the functions \(f\) in
Equation~\eqref{eq:base-model} \parencite{Stone1985}. These functions are linear
at the endpoints and require fewer parameters than classic cubic splines. We
call $k$ the number of knots used to define restricted spline functions. We
identify three cases according to the desired level of smoothness for each
variable. Theoretically, it is possible to set any arbitrary positive integer
value of \(k\) having the form,

\begin{equation*}
	f(X)= \beta_{1} X +  \sum_{l=2}^{k}  \beta_{l} S_{l}^{(k)}(X)
\end{equation*}

for $l=2, \ldots, k-2$,

\begin{align*}
	S_{l}^{(k)}(X) & = \left(X-t_{l}\right)_{+}^{3}-\left(X-t_{k-1}\right)_{+}^{3}\left(t_{k}-t_{l}\right) /\left(t_{k}-t_{k-1}\right) \\
	               & +\left(X-t_{k}\right)_{+}^{3}\left(t_{k-1}-t_{l}\right) /\left(t_{k}-t_{k-1}\right)
\end{align*}

The model's complexity increases in the same direction as \(k\). To keep a sane
number of parameters, the recommendation is to use \(k\) equal to 3, 4 or 5. The
exact positions of the \(t_i\)'s are defined as
\(\mathbb{P}(X\leq t_{i}) = q_{i}\) with the \(q_{i}\)'s defined in
Table~\ref{tab:knots-splines} for a given number of knots \(k\).

\begin{table}
	\centering
	\begin{tabular}{clllll}
		\toprule
		  & \multicolumn{5}{c}{Quantile level (\(q_i\))}                                         \\
		\cmidrule(lr){2-6}
		\(k\)
		  & \(q_1\)                                      & \(q_2\) & \(q_3\) & \(q_4\) & \(q_5\) \\
		\midrule
		3 & 0.1                                          & 0.5     & 0.9                         \\
		4 & 0.05                                         & 0.35    & 0.65    & 0.95              \\
		5 & 0.05                                         & 0.275   & 0.5     & 0.725   & 0.95    \\
		\bottomrule
	\end{tabular}
	\caption{Quantile levels (\(q_i\)) defining the knot  positions for cubic restricted spline functions.}
	\label{tab:knots-splines}
\end{table}

\section{Methodology}
\label{sec:methodology}

In this section, we will explain a series of steps to determine the most relevant variables in a model combining the classic relative weight analysis, nonlinear smoother with restricted cubic spline and residualization of interactions.  We can summarize our procedure in three steps:

\begin{enumerate}
    \item Select the best submodel with given a set of control, free, fixed and interaction terms. 
    \item Residualize all the interaction terms to remove the effects of the main variables and keep only the pure interaction effect. 
    \item Apply the relative weight analysis to detect which variables are the most significant.
\end{enumerate}

The steps are explained thoroughly in the following sections. 

\subsection{Model selection with interactions}

Given a model (Gaussian or logistic) and once defined the structure of the
variables (linear or nonlinear), we will define the interactions pairwise
between the variables. We will use  restricted interaction multiplication.

The restricted interaction will remove doubly nonlinear terms, allow us to
remove no essential terms. For example, assume we decide to model $A$ and $B$
with a 3 knot spline. The interaction between $A$ and $B$ is denoted by
\begin{equation*}
	f_{A}(A) \times f_{B}(B) =\beta_{1}\ A B + \beta_{2}\  A S_{1}^{(3)}(B)+\beta _{3}\   S_{1}^{(3)}(A)  B,
\end{equation*}

for some constants \(\beta_{1}\), \(\beta _{2}\) and \(\beta _{3}\). A similar
pattern follows the 4 knots case.

Setting a full set of interaction with \(p\) variables, we should fit
\(\frac{1}{2}(p-1)(p)\) interactions. For a model with k-knots spline functions,
we will need to fit \(1 + kp+ \frac{k}{2}(p-1)(p)\) distinct parameters. For
example, if we have \(p = 10\) variables, for a full interaction model with a
3-knots case, we will need to fit \(166\) parameters. If we decide to use the
4-knots case, the number increases to \(221\). Any model with such many
parameters is inadequate. The overfitting leads to erroneous results,
especially on small samples.

To solve this issue, we opted by selecting a submodel from the full model. We
use the classic stepwise method based on the change of Bayesian Information
Criterion (BIC). Recall that
\(\mathrm{BIC} = -2\log(\mathcal{L}) + \ln(n)v\). Here \(\mathcal{L}\)
represents the log-likelihood of the submodel and \(v\) the variables used to
fit it. In the procedure, we search to minimize the \(\mathrm{BIC}\). The factor
\(\ln(n)v\) strongly penalize large models unless the value
\(-2\log(\mathcal{L})\) is less than \(2v\).

In our context, the \(\mathrm{BIC}\) presents some advantages over the Akaike
Information Criterion (AIC). The AIC is estimated with
\(\mathrm{AIC} = -2\log(\mathcal{L}) + 2v\). The criterion penalizes the
submodel to exclude unnecessary variables as well as BIC using the factor
\(2v\). However, it smooths the selection, allowing more variables inside the
final model. Such models are appropriate for prediction instead to be
parsimonious \parencite[see][]{DziakSensitivity2020}. We focus on the method in the
inference of the most relevant features of the data, considering the BIC a
better value for the stepwise selection process.

In our implementation, we start with a model using only main effects. Then, we
continue adding or removing main effects or interactions until the algorithm
cannot improve the BIC value. The implementation of  the procedure was taken
from the package \texttt{MASS} \parencite{VenablesModern2002} using the function \texttt{stepAIC}. To use the
BIC, as mentioned before, we set the parameter \texttt{k = log(n)}. The function
allows to include a parameter \texttt{scope} that consists of a list with lower
and upper models. Given the structure of Equation~\eqref{eq:base-model}, we
define the lower and upper models as,
\begin{align*}
	\text{Lower Model} & =  \beta _{0} + \sum_{i=1} ^{c} \gamma_{i} Q_{i}  +  \sum_{i=1}^{d_{F}} f_{i}(X_{i}) \\
	\text{Upper Model} & = 	\beta_{0} + \sum_{i=1}^{c} \gamma_{i}Q_{i} + \sum_{i=1}^{d_F} f_{i}(X_{i})  \\
	& \quad \qquad + \sum_{i=d_F+1}^{d} f_{i}(X_{i}) + \sum_{i=1}^{d} \sum_{j=1}^{d} f_{i}(X_{i}) \times f_{j}(X_{j})
\end{align*}

The implementation include in the lower model only the control and fixed
variables. The upper model contains all the main effects and all the possible
interactions.




After selecting the most relevant variables on the model we remain with the
subsets \(f_{i_{1}}, \dots , f_{i_{d_{M}}}\) of main effects and
\(f_{i_{1}}(X_{i_{1}}) \times f_{j_1}(X_{j_{1}}), \dots f_{i_{d_{I}}}(X_{i_{d_{I}}}) \times f_{j_{d_{I}}}(X_{j_{d_{I}}})\)
of interactions. Here \(d_{M}\) and \(d_{I}\) represent the number of main
effects and interactions selected and \(\{i_{s}\}_{s=1}^{d_{M}}\) and
\(\{(i_{s},j_{s})\}_{s=1}^{d_{I}}\) are their respective indices.

The final model after the stepwise selection is
\begin{equation}
	\label{eq:final-model}
	\mathrm{C}(Y\vert \boldsymbol{X})  =
	\beta_{0} + \sum_{i=1}^{c} \gamma_{i}Q_{i}
	+ \sum_{s=1}^{d_{M}} f_{i_{s}}(X_{i_{s}})
	+ \sum_{s=1}^{d_{I}} f_{i_{s}}(X_{i_{s}}) \times f_{j_{s}}(X_{j_{s}}).
\end{equation}

Note finally that \(d_{M} \geq d_{F}\) and \(d_{I}\geq 0\). If \(d_{M}=d_{F}\)
then the model has only main effects. If \(d_{I}=0\) then none of the
interactions were added. For every interaction added, the main effects are also
added automatically to preserve the hierarchical principle.

\subsection{Residualized relative importance}

Interactions play a key role in this work. We want to know if each interaction
included, adds relevant information to the model. Otherwise, it is
negligible compared to its main effects. Here, the interaction contains
little information blurring other results in the model. The objective is to
separate the relevant effects from the main ones and interactions.

Notice that in the case of simple linear interactions like \(X_i \times X_j\), there
are three effect types

\begin{enumerate}
	\item The effect solely from \(X_i\).
	\item The effect solely from \(X_j\).
	\item The effect solely from the interaction of \(X_i \times X_j\).
\end{enumerate}

Interactions with restricted splines, are handle it equally. Except that the
effects are more diffused across the terms. For example, for a 3 knot spline,
the interaction between \(X_i\) and \(X_j\) is

\[
	f_i(X_i) \times f_j(X_j) = \beta_1 X_i  X_j + \beta_2 X_i S_{1}^{(3)}(X_j) + \beta_3 S_{1}^{(3)}(X_{_{i}}) X_j
\]

The three terms contain mixed information about \(X_i\) and \(X_{j}\).
Therefore, if we apply the relative weight analysis to this interaction, the
different effects are blurred. By controlling first by the other variables, we
isolate the pure interaction effect.

Suppose that we have a simpler model, where \(f_1\) and \(f_2\) are 3-knots
restricted splines,
\begin{equation*}
	\mathrm{C}(Y\vert (X_1,X_2))  = \beta_{0} + f_{1}(X_{1}) +f_{2}(X_{2})  + f_{1}(X_{1})\times f_{2}(X_{2})+ \epsilon
\end{equation*}

An extended way to state this model is also

\begin{align*}
	\mathrm{Logit}(\mathbb{P}(Y=1))
	 & = \beta_{0}                                                                           \\
	 & + \beta _{1} X_1 + \beta _{2} S_{1}^{(3)}(X_1)                                        \\
	 & + \beta_{3} X_2 + \beta_{4} S_{1}^{(3)}(X_2)                                          \\
	 & + \beta_5 X_1  X_2 + \beta_6 X_1 S_{1}^{(3)}(X_2) + \beta_7 S_{1}^{(3)}(X_{_{1}}) X_2 \\
	 & + \epsilon
\end{align*}

The procedure proposed in \textcite{LeBreton2013} is the following:

\begin{enumerate}
	\item Replace the higher order term with the residual obtained after regress
	      the interaction with respect to their main effects.
	      \begin{align*}
		      X_{1}X_{2}               & = \tilde{\beta}_{1} X_{1} + \tilde{\beta}_{2} X_{2} +r_{X_1 X_2}                           \\
		      X_1 S_{1}^{(3)}(X_2)     & = \tilde{\beta}_{3} X_{1} + \tilde{\beta}_{4} S_{1}^{(3)}(X_2)  +r_{X_1 S_{1}^{(3)}(X_2) } \\
		      S_{1}^{(3)}(X_{1}) X_{2} & = \tilde{\beta}_{5} S_{1}^{(3)}(X_{1}) + \tilde{\beta}_{6} X_2  +r_{S_{1}^{(3)}(X_1) X_2}  \\
	      \end{align*}
	      Here the variables \(r\) represent the residual of the regressions. We
	      perform the regression without an intercept to capture the full effects
	      from the residual \(r_{12}\).
	\item Refit the model with the form
	      \begin{align}
		      \begin{split}
			      \label{eq:design-residualization}
			      \mathrm{C}(Y\vert (X_{1},X_{2}))
			      &= \beta_{0} \\
			      & + \beta _{1} X_1 + \beta _{2} S_{1}^{(3)}(X_1)     \\
			      & + \beta_{3} X_2 + \beta_{4} S_{1}^{(3)}(X_2)  \\
			      & + \beta_5 r_{X_1  X_2} + \beta_6 r_{X_1 S_{1}^{(3)}(X_2)} + \beta_7 r_{S_{1}^{(3)}(X_{_{1}}) X_2} .
		      \end{split}
	      \end{align}

	      It means that the interactions are replaced by the residuals of the step
	      before. We denote the residualized interaction term as
	      \[
		      \widetilde{f}_{1}(X_1)\times \widetilde{f}_2(X_2) = \beta_5 r_{X_1 X_2} + \beta_6 r_{X_1 S_{1}^{(3)}(X_2)} + \beta_7 r_{S_{1}^{(3)}(X_{_{1}}) X_2}
	      \]
\end{enumerate}

The main advantage of residualizing the interaction effects, we create a
new set of interaction variables uncorrelated with their main effects.
The procedure separates the true synergy between the variables and the
main effects. However, the effects could be correlated with each other.

We apply this algorithm to all the restricted interactions of the reduced
model. Therefore, the RWA was applied for the interactions over the
residuals of an intermediate linear regression between the main effect.

\subsection{Relative Weight Analysis}
\label{sec:org5aee3e4}

Relative Weight Analysis is tool used to separate the relevance of each
variable or interaction. This technique creates a new set of predictors with are
an orthogonal representation of the observed predictors (with one-to-one
correspondence). Estimate the influence on this new space and then return the
results to the original space. This way the problem presented with correlated
models disappears given that the importance estimation occurs in this new
space.

We are interested in the importance of the main effects versus the interactions.
The design matrix can have multiple patterns. For example, the right side of
equation \eqref{eq:design-residualization}. Here a mix of linear and nonlinear
variables is present. We are interested in determining the effect of the whole
variables instead of their components. Therefore, we will define the matrix
\(X\) as
\begin{align*}
	\boldsymbol{D}
	 & = \bigl[ Q_{1}, \dots , Q_{c},                                                 \\
	 & \qquad f_{i_1}(X_{i_1}), \dots , f_{i_{d_{M}}}(X_{i_{d_{M}}}),                 \\
	 & \qquad \widetilde{f}_{i_1}(X_{i_{1}})\times \widetilde{f}_{j_1}(X_{j_1}),\dots
	\widetilde{f}_{i_{d_{I}}}(X_{i_{d_{I}}})\times \widetilde{f}_{j_{d_{I}}}(X_{j_{d_{I}}})   \bigr].
\end{align*}

Therefore, the matrix \(\boldsymbol{D}\) contains the control variables and all
the effects and interactions resulting from the selection model procedure. Also,
the interactions contain only information from the pure synergy between the
variables due to the residualization.

The algorithm to estimate the relative weights is the following:

\begin{enumerate}

	\item Start with design matrix \(\boldsymbol{D}\).

	\item Standardize the columns of \(\boldsymbol{D}\).
	\item Estimate the singular value decomposition
	      \(\boldsymbol{D} = A \Delta B ^{\top}\), where

	      \begin{enumerate}
		      \item \(A\) is the matrix of eigenvectors associated to
		            \(\boldsymbol{D} \boldsymbol{D} ^{\top}\).

		      \item \(B\) is the matrix of eigenvectors associated to
		            \( \boldsymbol{D}^{\top}\boldsymbol{D} \).

		      \item \(\Delta\) is the diagonal matrix of singular values, which is
		            equivalent to the square root of eigenvalues of
		            \(\boldsymbol{D}^{\top}\boldsymbol{D} \).

	      \end{enumerate}

	\item Create the orthogonal version \(Z = AB^{\top}\) of \(D\) using the SVD
	      decomposition. The columns \(Z\) are one-to-one orthogonal
	      representations of the columns of \(\boldsymbol{D}\).
	\item Standardize the columns of \(Z\).

	\item Estimate the fully standardized coefficient for the model
	      \(C(Y\vert Z) = \beta Z\). However, depending on the link function the
	      steps differ.

	      \begin{description}
		      \item[Gaussian:] \(C(Y\vert \boldsymbol{X}) = Y\).

			      \begin{enumerate}
				      \item Estimate the standardized ordinary least square
				            coefficient of \(Z\) with the formula
				            \[\beta^{*}=(Z^{\top}Z)^{-1}Z^{\top}Y\].
			      \end{enumerate}

		      \item[Logistic:] \(\mathrm{C}(Y\vert \boldsymbol{X}) = \mathrm{logit}(\mathbb{P}\left(Y=1\vert \boldsymbol{X} \right))\).

			      \begin{enumerate}
				      \item Obtain the unstandardized coefficients \(b\) of the
				            model against the matrix \(Z\). Here the \(b\) could
				            be calculated using a least square procedure.

				      \item Compute \(R_{L}^{2}\) which is.
				            \begin{equation*}
					            R^2_{L} = 1- \frac{\sum (Y-\hat{Y}) ^{2}}{\sum (Y-\bar{Y})^{2}}
				            \end{equation*}
				            Alternatively, regress \(Y\) against \(\hat{Y}\) and recover
										the \(R^{2}\).
				      \item Estimate the standard deviation of the logit
				            prediction, \(s_{\mathrm{logit{(\hat{Y})}}}\), where
				            \(\operatorname{logit}(\hat{Y}),=\ln [\hat{Y} /(1-\hat{Y})]\)
				      \item Estimate the fully standardized coefficients
				            \[
				            \beta^{*}=(b)\left(s_{Z}\right)\left(R_{L}\right) /\left(s_{\mathrm{logit}(\hat{Y})}\right)
				            \]
				            Here \(s_{Z}\) is the standard deviation of each
				            column of \(Z\).
			      \end{enumerate}

	      \end{description}
	\item  Create the link transformation \(\Lambda^{*}=(Z^{'}Z)^{-1}Z^{'}X\).
	\item Estimate the relative weights with
	      \[
		      \epsilon = {\Lambda^{*}}^{2} {\beta^{*}}^{2}
	      \]
\end{enumerate}

The procedure projects the original columns of \(\boldsymbol{D}\) to orthonormal
space called \(Z\) with a correspondence one-to-one in their columns. In this
new space, the fully standardized coefficients of the model
\(C(Y\vert Z) = \beta Z\). Those are named as \({\beta ^{*}}^{2}\). A particular
property is \({\beta ^{*}}^{2}\) are the relative importance of the columns in
\(Z\). Given that \(Z\) is orthonormal, the weights refer to a uncorrelated
variables set. To return those weights onto the original dimension space, we
estimate a transformation matrix called \(\Lambda\). The matrix \(\Lambda\)
transforms the \({\beta^{*}}^{2}\) into the original scale \(\boldsymbol{D}\).
The sum of the \(\epsilon \) are equal to the \(R^2_{O}\) or \(R_{L}^{2}\)
depending on the case. Finally, each variable has associated a percentage
contribution to the explained variance.

\section{Results}
\label{sec:results}

We will show the results of our algorithm. To test the
algorithm, we will generate a sample of \(n=3000\) of random variables
determined by each example. In the simulated examples, we tested the capability
of each method to determine the most relevant depending on if we are in the
Gaussian or binomial case.

We follow these steps to generate two outputs variable \(Y_{g}\) and their
respective dichotomized version \(Y_{b}\), with a model with \(d\) input
variables \(X_1, \dots , X_{d}\):

\begin{enumerate}
	\item Generate a variable \(Y_{g}\) with the functional form of the variable. For
	      example, \(Y_{g}= X_1 + X_2\).
	\item Apply the inverse of the logistic function,
	      \(p(L) = \mathrm{logit}^{-1}(Y_{g}) \).
	\item Generate a Bernoulli random variable ($\{0,1\}$) with probability
	      outcome \(p(Y_{g})\).
	\item Repeat this procedure for each element of the sample. Call this vector
	      \(Y_{b}\).
\end{enumerate}

The described step will generate a random variable Bernoulli \(Y_{b}\) with values
\(\{0,1\}\) and generate with the underlying model given by \(Y_{g}\).

For testing, we will use three classic models: The Ishigami model, the
\(g\)-model, and a syntethic model of 15 variables by
\textcite{OakleyProbabilistic2004}.

\subsection{Ishigami function}

The Ishigami function is classic to test the sensitivity and reliability variables
in models. It presents a strong non-linearity and non-monotonicity. The order of
the variables in terms of relevance are \(X_2\), \(X_1\) and the interaction
\(X_1\) with \(X_3\). All other variables and interaction have zero
theoretical relevance's. The form of the model is

\begin{equation*}
	Y_{g}= \sin X_1 + 7\ \sin^2 X_2 + 0.1\ X_3^4 \sin X_1
\end{equation*}
where $X_i\sim \mathrm{Uniform}(-\pi, \pi)$ for $i=1,2,3$, $a = 7$ and
$b = 0.1$. We also set \(X_4 - X_8\sim \mathrm{Uniform}(-\pi, \pi)\) to add
uncorrelated noise to the model.

In the work of \textcite{IshigamiImportance1990}, they split the model's variance
to determine the relevance of each predictor. They concluded in this model that
the percentage of variance due to \(X_1, X_{2}\) and the interaction \(X_{13}\)
are \(31.38\%\), \(44.24\%\) and \( 24.36\%\) respectively. The other effects
have 0\% on their contribution to the variance.

We applied our procedure using the \(Y_{g}\) and \(Y_{b}\) cases.
Table~\ref{tab:ishigami} presents the results for both cases. Notice how the
variables \(X_2\), \(X_1\) and the interaction, \(X_1 \times X_3\) are captured
correctly. For the three mentioned effects, we choose a restricted cubic spline
with 5-knots. Recall that the model tested all the main effects and cross terms
(\(X_1\times X_2\), \(X_{1}\times X_3\), \(X_1\times X_4\), and so on). However,
as we are using the BIC as a selection criterion, all the non-relevant terms were
removed by the large penalty \(\ln(n)\) in the criterion.

In terms of goodness-of-fit, the \(R^2_{O}=0.8308\) while \(R^2_{L} = 0.4654\).
The results are expected. The Ishigami model is continuous and
from it, we create two outputs \(Y_{g}\) and \(Y_{b}\). In the binary case we
loss power to predict the variance given the change of the dependent variable.
Even in the case, active effects were detected.

\begin{table}[ht]
	\centering
	\begin{tabular}{llrl}
		\toprule
		 & Variable          & Weight  & Type         \\
		\midrule
		Continuous output \(Y_{g}\)
		 & $X_1$             & 34.52\% & Free         \\
		 & $X_2$             & 48.27\% & Free         \\
		 & $X_3$             & 0.09\%  & Free         \\
		 & $X_1 \times  X_3$ & 17.12\% & Interactions \\
		\midrule
		Binary output \(Y_{b}\)
		 & $X_1$             & 31.33\% & Free         \\
		 & $X_2$             & 53.25\% & Free         \\
		 & $X_3$             & 2.90\%  & Free         \\
		 & $X_1\times  X_3$  & 12.52\% & Interactions \\
		\bottomrule
	\end{tabular}
	\caption{Relative weights for the Ishigami model when using the continuous
		output \(Y_{g}\) and binary output \(Y_{b}\). We use  5-knots
		restricted cubic splines as smoother. }
	\label{tab:ishigami}
\end{table}

To present the capabilities of the algorithm, we set two disfavorable
scenarios for the Ishigami model. In these settings we will use only \(Y_{g}\).

First, we set as fixed the variables $X_4$, $X_5$, $X_6$, $X_7$ and $X_8$. This setting
will cause that the model is forced to fit 5 noisy variables with little
information for \(Y_{g}\). The model fit all the possible interactions, and
removes those that worsen BIC. However, the variables \(X_4-X_8\) will
remain in all models.

The other scenario is when use \(X_3\) a control variable. Here, the
variable \(X_3\) is in all models, but none interaction contain
\(X_{3}\). This setting is useful when the study requires controlling by some
variables, but you don't want any further interference of it (e.g., gender, age,
smoke, etc.).

Table~\ref{tab:ishigami-fixed-control} presents the values on the mentioned
scenarios. In the upper part of the table we notice how the inclusion of the
non-important variables does not affect the weights compared with
Table~\ref{tab:ishigami}. In fact, all variables \(X_4-X_8\) present values
between 0.04\%-0.09\%. If needed, the analyst can remove variables with weights
in this order of magnitude without losing any inference power. The lower part
presents the case when we control \(X_3\) in the model. Notices how the model
relies on  the weights of \(X_1\) and \(X_2\) because we restricted the use of
the interaction \(X_1\times X_3\). In this case

\begin{table}[ht]
	\centering
	\begin{tabular}{llrl}
		\toprule
		 & Variable        & Weight  & Type        \\
		\midrule
		\multirow{2}{3cm}{Fixed  $X_4$, $X_5$, $X_6$, $X_7$ and $X_8$}
		 & $X_1$           & 35.57\% & Free        \\
		 & $X_2$           & 48.12\% & Free        \\
		 & $X_3$           & 0.06\%  & Free        \\
		 & $X_4$           & 0.06\%  & Fixed       \\
		 & $X_5$           & 0.05\%  & Fixed       \\
		 & $X_6$           & 0.04\%  & Fixed       \\
		 & $X_7$           & 0.09\%  & Fixed       \\
		 & $X_8$           & 0.05\%  & Fixed       \\
		 & $X_1\times X_3$ & 15.96\% & Interaction \\
		\midrule

		Controlled \(X_3\)
		 & $X_1$           & 42.69\% & Free        \\
		 & $X_2$           & 57.16\% & Free        \\
		 & $X_3$           & 0.15\%  & Control     \\
		\bottomrule
	\end{tabular}
	\caption{Relative weights when fixing the variables \(X_4, X_{5}, X_6, X_7\)
		and \(X_8\) (upper) and when controlling by \(X_{3}\) (lower).
		All the results were estimated with the continuous output \(Y_{g}\)}
	\label{tab:ishigami-fixed-control}
\end{table}


\subsection{The Moon function}
\label{sec:moon-function}

The work of~\textcite{MoonDesign2010,MoonTwoStage2012} established an algorithm
to detect active and inactive variables on complex configurations. They propose
a 20-dimensional function with 5 active main effects $X_1$, $X_7$, $X_{12}$,
$X_{18}$, $X_{19}$ and 4 active effects \(X_1\times X_{18}\),
\(X_1 \times X_{19}\) \(X_7 \times X_{12}\) and the quadratic term \(X_{19}^2\).

Assuming  \(X_{i}\sim Unif(0,1),\ i=1 , \dots , 20\), the explicit form of the model is,
\begin{equation*}
	Y_{g}^\text{base} = -19.71 X_{1} X_{18}+23.72 X_{1} X_{19}-13.34 + X_{19}^{2}+28.99 X_{7} X_{12}+ \text{small terms}.
\end{equation*}

They studied another complex configuration, including or removing the small terms
and amplification of the active effects. To test if the procedure detects the
relevant variables, we use the version where all the active effects were tripled
while the small terms remain constant,
\begin{equation*}
	Y_{g}^{C_3}=-59.13 X_{1} X_{18}+71.16 X_{1} X_{19}-40.02 X_{19}^{2}+86.97 X_{7} X_{12} + \text{small terms}.
\end{equation*}

The \textit{small terms} part consist of 189 elements between main, quadratic
and interaction terms with low impact on the \(Y_{g}\). The coefficients can be
found in \textcite{MoonDesign2010} (Table 3.11). The aim setting
\(Y_{g}^{\text{base}}\) and \(Y_{g}^{C_{3}}\) is to compare the changes on the
relative weights in  disfavorable and favorable scenarios.

Table~\ref{tab:moon-yg} presents the result using our algorithm with
\(Y_{g}^\text{base} \) and $Y_{g}^C{_3}$. We estimated the absolute difference
between both values to compare their change. The relative weights for
\(Y_{g}^\text{base} \) give significantly weight to non-important variables like
\(X_3\) or \(X_{15}\). The behavior is due to the variable \(Y_{g}^\text{base}\)
presents a more blurred information about the true active effects. However, the
effects $X_1\times X_{18}$, $X_1\times X_{19}$, $X_7\times X_{12}$ are well
represented in the model. s

With $Y_{g}^{C_3}$ all the non-important variables decrease their value. The
true active ones increase their participation greatly. The variables \(X_7\),
\(X_{12}\), \(X_{18}\) and $X_7 \times X_{12}$ increase more than 10 percentage
points with respect to the other model. Other terms like $X_1$, $X_{18}$,
$X_1\times X_{18}$ and $X_1\times X_{19}$, increase in a small amount.

The case for the quadratic term with $X_{19}$ is interesting. Its value is low
in both scenarios. However, recall that we are using a restricted cubic spline
with 5 knots to fit the mains and interaction effects. Additionally, given the
residualization step, most of the participation of $X_{19}$ could be
represented through $X_1\times X_{19}$ the residualization process to extract
all the effects from the interactions to the main present lower, the effect

We observe how the residual weights are more predominant for the $C_3$ case than
the base. The result is expected because small terms are stronger in
the base case than in the $C_3$.

\begin{table}[ht]
	\centering
	\begin{tabular}{lrrr}
		\toprule
		                                     &
		\multicolumn{2}{c}{Relative weights} &                                     \\
		\cmidrule(lr){2-3}

		Variable                             &
		\(Y_{g}^{\text{base}}\)              &
		\(Y_{g}^{C_{3}}\)                    &
		\(\Delta\) (p.p.)                                                          \\
		\midrule
		$X_{1}$                              & 1.26\%  & 1.38\%  & \textbf{0.13}            \\
		$X_{2}$                              & 1.21\%  & 0.13\%  & -1.08           \\
		$X_{3}$                              & 16.30\% & 2.97\%  & -13.33          \\
		$X_{4}$                              & 2.51\%  & 0.36\%  & -2.15           \\
		$X_{5}$                              & 3.24\%  & 0.56\%  & -2.68           \\
		$X_{6}$                              & 0.49\%  & 0.05\%  & -0.44           \\
		$X_{7}$                              & 1.25\%  & 12.21\% & \textbf{10.97}  \\
		$X_{8}$                              & 1.40\%  & 0.22\%  & -1.17           \\
		$X_{9}$                              & 1.97\%  & 0.26\%  & -1.71           \\
		$X_{12}$                             & 2.20\%  & 14.29\% & \textbf{12.10 } \\
		$X_{13}$                             & 0.95\%  & 0.13\%  & -0.81           \\
		$X_{14}$                             & 3.27\%  & 0.71\%  & -2.57           \\
		$X_{15}$                             & 11.71\% & 1.69\%  & -10.02          \\
		$X_{16}$                             & 5.05\%  & 0.95\%  & -4.10           \\
		$X_{17}$                             & 4.25\%  & 0.56\%  & -3.69           \\
		$X_{19}$                             & 0.83\%  & 2.62\%  & \textbf{1.79}            \\
		$X_{20}$                             & 3.25\%  & 0.51\%  & -2.74           \\
		$X_{18}$                             & 10.19\% & 11.42\% & \textbf{1.24}   \\
		$X_{1} \times X_{18}$                & 9.00\%  & 10.18\% & \textbf{1.18}   \\
		$X_{1} \times X_{19}$                & 6.77\%  & 8.32\%  & \textbf{1.56}   \\
		$X_{7} \times X_{12}$                & 12.92\% & 30.46\% & \textbf{17.55}  \\
		\bottomrule
	\end{tabular}
	\caption{Relative weights for the Moon model using the variable
		\(Y_{g}^{\text{base}}\) and \(Y_{g}^{C_{3}}\). The absolute difference
		between both weights (\(\Delta\)) is estimated in percentage points (p.p.).
		We use 5-knots restricted cubic splines as smoother.}
\label{tab:moon-yg}
\end{table}

\section{Conclusion}
\label{sec:conclusion}

In this paper, we explore a method to detect relevant variables using the
relative weight analysis technique. %
The main contribution in our algorithm is the residualization of the
interactions to capture the true effect of them. %
In a classic setting, the relative weights for the main effects and interactions
are evaluated as is, without considering the real nature of the latter. %
In other words the interaction of two variables contains information on each
variable separately and the remaining belongs to the true interaction effect. %
In this work, we emphasize this technique following the works
of~\textcite{JohnsonHeuristic2000,TonidandelDetermining2010,LeBreton2013}. %

One aim was to create a flexible algorithm beyond to the classic linear model.
To this end, we include the restricted cubic spline as smothers to determine the
relationship between the inputs and outputs. %
This feature allows us to detect nonlinear patterns in the data, even in complex
settings. %
More work should be done to make the procedure flexible to a wide set of cases. %

We can find packages \texttt{R} performing Residual Weight Analysis like
\texttt{rwa} \parencite{ChanRwa2020} or \texttt{flipRegression}
\parencite{DisplayrFlipRegression2021}. They rely on the analysis of main
effects. %
In this study, we implemented a version that allows the user to set control,
fixed, free and interactions in model. Besides, the predictors are modeled using
restricted spline smothers. In this context, the analysis should be more
accurate specially if the phenomenon is highly nonlinear. The great advantage
with respect to other techniques is the analysts can include some variable in
the model even if they are non-important. %
Or they can compare the weight between main effects on interactions to choose a
particular model. %

The stepwise procedure produced effective results, even if they are some points
against it  \parencite[e.g.][]{MillerSubset2002,HarrellRegression2015}. %
Other techniques to model selection were also considered like elastic nets
\parencite{ZouRegularization2005} or PLS regression
\parencite{MartensReliable2001,MevikPls2007}. %
They represent interesting lines of research to follow in the future. %
In this study, we considered the implementation because of its simplicity. %
Stepwise regression combined with the Bayesian Information Criterion (BIC),
is a fast and effective method to create the most relevant and simpler model. %
The BIC cut all the non-relevant variables in the first steps allowing the
model to include those interactions adding some real value to the model. %
Other information criteria \parencite{DziakSensitivity2020} can be explored in
the future. %

Finally, even if the results establish that our procedure identify correctly the
relevant parts of our models, a proper validation must be done to check the
dispersion of the relative weights. %
If we set a configuration over different realizations, the questions remaining
are if the selected variables keep constant and how much the relative weights
differ from each other. %
A proper validation of the results are beyond the scope of this paper, but, the
work of \cite{TonidandelDetermining2009} conducted an extensive study of simulation
with bootstrap confidence intervals. %
The implementation of the technique will be a priority for future
developments. %

\FloatBarrier

\section*{Declarations}
\subsection*{Funding}

The authors acknowledge the financial support from Escuela de Matemática de la
Universidad de Costa Rica, through CIMPA, Centro de Investigaciones en
Matemática Pura y Aplicada through the projects 821–B8-A25.

\subsection*{Code availability}

All the calculations in this package were made using an own R-package
residualrwa. The package is published on github on this address
\url{https://github.com/maikol-solis/residualrwa}.

\subsection*{Conflict of interest}

On behalf of all authors, the corresponding author states that there is no
conflict of interest.

\printbibliography




\end{document}